\documentclass[journal]{IEEEtran}
\ifCLASSINFOpdf
\else
   \usepackage[dvips]{graphicx}
\fi

\usepackage{lipsum} 
\usepackage{url}
\usepackage{graphicx}
\usepackage{cite}
\usepackage{balance}
\usepackage{hyperref}
\usepackage{url}
\usepackage{amsmath,amssymb,amsfonts}
\usepackage{algorithmic}
\usepackage{graphicx}
\usepackage{textcomp}
\usepackage{xcolor}
\usepackage{amsthm}
\usepackage{subcaption}
\usepackage{balance}
\usepackage{mathrsfs}
\usepackage{subcaption}
\usepackage{algorithm}
\usepackage{algorithmic}
\usepackage{authblk}

\newtheorem{proposition}{Proposition}
\newtheorem{lemma}{Lemma}

\begin{document}

\title{Quantile Learn-Then-Test: Quantile-Based Risk Control for Hyperparameter Optimization}

\author{Amirmohammad Farzaneh, \IEEEmembership{Member, IEEE}, Sangwoo Park \IEEEmembership{Member, IEEE}, and Osvaldo Simeone, \IEEEmembership{Fellow, IEEE}\thanks{This work was supported by the European Union’s Horizon Europe project CENTRIC (101096379), by the Open Fellowships of the EPSRC (EP/W024101/1), and by the EPSRC project (EP/X011852/1). The authors are with the King’s Communications, Learning \& Information Processing (KCLIP) lab within the Centre for Intelligent Information Processing Systems (CIIPS), Department of Engineering, King’s College London, WC2R 2LS London, U.K. (e-mail: amirmohammad.farzaneh@kcl.ac.uk; sangwoo.park@kcl.ac.uk; osvaldo.simeone@kcl.ac.uk)}}

\maketitle

\begin{abstract}
    The increasing adoption of Artificial Intelligence (AI) in engineering problems calls for the development of calibration methods capable of offering robust statistical reliability guarantees. The calibration of black box AI models is carried out via the optimization of hyperparameters dictating architecture, optimization, and/or inference configuration. Prior work has introduced learn-then-test (LTT), a calibration procedure for hyperparameter optimization (HPO) that provides statistical guarantees on average performance measures. Recognizing the importance of controlling risk-aware objectives in engineering contexts, this work introduces a variant of LTT that is designed to provide statistical guarantees on quantiles of a risk measure. We illustrate the practical advantages of this approach by applying the proposed algorithm to a radio access scheduling problem.
\end{abstract}


\IEEEpeerreviewmaketitle


\section{Introduction}

\subsection{Motivation and Overview}
\label{sec:motivation}

Artificial Intelligence (AI) is increasingly adopted in engineering domains, such as wireless systems \cite{letaief2019roadmap}, as a tool to facilitate design and to reduce implementation complexity. In practice, the effectiveness of AI models relies on the selection of hyperparameters such as architectural aspects including the number of layers of a neural network, optimization parameters like the learning rate for fine tuning, or inference parameters such as the thresholds used for multi-label classification models \cite{wu2019hyperparameter, feurer2019hyperparameter}. Accordingly, hyperparameter optimization (HPO) methods, which seek to identify optimal hyperparameters, have been widely investigated, with techniques ranging from random search \cite{bergstra2012random}, to Bayesian optimization \cite{wu2019hyperparameter} and bandit-based methods \cite{li2018hyperband} (see review papers \cite{bischl2023hyperparameter, feurer2019hyperparameter}). 

Conventional HPO methods do not provide guarantees on the test risk obtained with an optimized choice of hyperparameters. To fill this gap, the recently introduced \textit{learn-then-test} (LTT) method \cite{angelopoulos2021learn} provides statistical bounds on the \textit{average} risk of the returned hyperparameters by adopting a multiple hypothesis testing methodology \cite{rice2007mathematical}. Existing applications of LTT include medical imaging \cite{lu2022improving} and language modelling \cite{schuster2022confident}.

While LTT only offers guarantees on the average risk, system designers in the engineering domain typically wish to control more robust functionals of the risk such as \textit{quantiles}. Specifically, the designer is often interested in guaranteeing that a sufficiently large fraction of the realized performance measures meets a given target level. For example, in cellular wireless systems, the network operator aims at ensuring that \textit{key performance indicators} (KPIs), such as the throughput and latency, satisfy reliability requirements for a sufficiently large fraction of the users and/or of the time \cite{ganewattha2020real, valkova2017novel, cheng2015energy}.

With this motivation in mind, this letter builds on LTT to introduce a novel HPO methodology that is able to provide statistical guarantees on any desired quantile of the risk. 

\subsection{Related Work}

Conformal prediction algorithms offer statistical guarantees on the outputs of AI models by controlling thresholds used to produce prediction sets \cite{angelopoulos2023conformal}. For example, risk-controlling prediction sets (RCPS) ensure control over the average risk, with the option to extend this control to other risk functionals \cite{bates2021distribution}. The use of conformal prediction to manage quantiles of the risk has been explored in \cite{snell2022quantile}, while other risk functionals addressed by conformal methods include conditional value-at-risk (CVaR) \cite{snell2022quantile} and dispersion measures \cite{deng2024distribution}. While conformal prediction methods are limited to HPO problems involving simple prediction thresholds, LTT manages the average risk by optimizing over arbitrary hyperparameter vectors.

\subsection{Main Contributions}

This letter presents the following main contributions.

\begin{itemize}
    \item \textbf{Methodology: }We introduce a novel variant of LTT \cite{angelopoulos2021learn}, named QLTT, that provides hyperparameter sets with statistical guarantees on quantiles of the risk, extending the applicability of LTT beyond average risk control.

    \item \textbf{Applications:} We illustrate the practical application of QLTT by testing it on a popular radio access scheduling problem in communication engineering \cite{de2020radio}. This case study highlights the effectiveness of our approach in addressing complex real-world challenges, showcasing its capability in controlling risk quantiles in real-world scenarios.

\end{itemize}

The rest of the letter is arranged as follows. We begin by providing a description of the LTT procedure and its guarantees in Section \ref{sec:LTT}. We then introduce QLTT and prove its statistical guarantees in Section \ref{sec:QLTT}. In Section \ref{sec:simulations}, we test the application of QLTT on a wireless resource allocation problem. Finally, we conclude the paper in Section \ref{sec:conclusion} and introduce potential future research directions.

\section{Learn Then Test}
\label{sec:LTT}
In this section, we provide an overview of LTT, an HPO method that provides statistical guarantees on the \textit{average} risk \cite{angelopoulos2021learn}. Specifically, LTT returns a set of hyperparameters that guarantee an average risk below a user-defined threshold with high probability.

Assume a pre-selected discrete and finite set $\Lambda$ of hyperparameters $\lambda$ dictating the performance of a machine learning model. Define a risk function $R(Z,\lambda)$ which scores hyperparameter $\lambda$ on test point $Z$. The risk function is negatively oriented, so that a smaller risk corresponds to a better-performing hypervector. We also assume that the risk is bounded as $0\leq R(Z,\lambda) \leq 1$. For example, the risk function may measure the prediction error in a supervised learning setting in which the test data $Z = (X,Y)$ includes covariates $X$ and label $Y$.

The goal of LTT is to control the average risk 
\begin{equation}
\label{Eq:LTT_risk}
    R(\lambda) = \mathbb{E}_{\mathcal{Z}}\left[R(Z, \lambda)\right],
\end{equation}
where the average is taken over the unknown data distribution $p_\mathcal{Z}$. A hyperparameter $\lambda$ is thus deemed to be reliable if we have the inequality $R(\lambda) \leq \alpha$ for some user-defined threshold $\alpha$ with $0\leq \alpha \leq 1$. LTT uses calibration data $\mathcal{Z} = \left\{Z_i\right\}_{i=1}^n$ with i.i.d. samples $Z_i\sim p_{\mathcal{Z}}$ to produce a subset $\hat{\Lambda} \subseteq \Lambda $ of hyperparameters that are deemed to be reliable with high probability. Formally, the subset $\hat{\Lambda}$ satisfies the condition
\begin{equation}
\label{eq:LTT}
    \mathrm{Pr}_{\mathcal{Z}}[R(\hat{\lambda}) \leq \alpha \; \text{for all}\; \hat{\lambda} \in \hat{\Lambda}]\geq 1-\delta,
\end{equation}
where $\delta$ is a user-specified outage probability with $0\leq \delta \leq 1$, and the probability is taken over the calibration data $\mathcal{Z}$. By (\ref{eq:LTT}), all hyperparameters $\hat{\lambda}$ in the selected set $\hat{\Lambda}$ are reliable with probability at least as large as $1-\delta$.

To achieve this goal, LTT considers a \textit{multiple hypothesis testing} (MHT) approach whereby the null hypotheses
\begin{equation}
\label{eq:hypothesis}
    \mathcal{H}_\lambda : R\left(\lambda\right) > \alpha
\end{equation}
are tested for all candidate hyperparameters $\lambda \in \Lambda$. This way, rejecting the null hypothesis $\mathcal{H}_\lambda$ is equivalent to deciding that hyperparameter $\lambda$ satisfies the reliability constraint $R(\lambda)\leq \alpha$.

To test each hypothesis $\mathcal{H}_\lambda$ in (\ref{eq:hypothesis}), LTT obtains an empirical estimate 
\begin{equation}
\label{eq:average}
    \hat{R}(\lambda) = \frac{1}{n}\sum_{i=1}^n R(Z_i,\lambda),
\end{equation}
based on the calibration data $\mathcal{Z}$. Then, a $p$-value is formed for each null hypothesis $\mathcal{H}_\lambda$ by using a concentration inequality. A $p$-value for the null hypothesis $\mathcal{H}_\lambda$ is a random variable $p_\lambda$ satisfying the condition $\mathrm{Pr}[p_\lambda \leq u]\leq u$ for all $0\leq u \leq 1$. For instance, using Hoeffding's inequality, we obtain the $p$-value \cite{angelopoulos2021learn}
\begin{equation}
\label{eq:pval}
    p_\lambda = e^{-2n(\alpha - \hat{R}(\lambda))^2_+}.
\end{equation}

Finally, after having obtained the set of $p$-values for all hyperparameters $\lambda \in \Lambda$, the set $\hat{\Lambda}$ is evaluated by using an algorithm $\mathcal{A}$ that controls the family-wise error rate (FWER) as
    \begin{equation}
    \label{eq:fwer}
        \hat{\Lambda} = \mathcal{A}(\{p_\lambda\}_{\lambda \in \Lambda}).
    \end{equation}
By definition, FWER-controlling mechanisms satisfy the requirement (\ref{eq:LTT}) \cite{rice2007mathematical}.

As the most basic FWER-controlling scheme, the Bonferroni correction outputs the set
\begin{equation}
    \hat{\Lambda} = \left\{\lambda \in \Lambda: p_\lambda < \frac{\delta}{|\Lambda|}\right\},
\end{equation}
where $|\Lambda|$ is the cardinality of the set $\Lambda$. However, Bonferroni's correction becomes increasingly conservative as the size of the initial candidate set $\Lambda$ grows large. To address this problem, \textit{fixed sequence testing} (FST) \cite{bauer1991multiple} can be used if there is some apriori knowledge about the hyperparameters that are more or less likely to be reliable. In this case, the candidate hyperparameters $\lambda \in \Lambda$ are preordered as $(\lambda_1,\ldots,\lambda_{|\Lambda|})$, where $\lambda_1$ and $\lambda_{|\Lambda|}$ are the hyperparameters that are expected to be the most and least reliable, respectively. FST considers each hyperparameter $\lambda_i$ in the order of decreasing expected reliability $\lambda_1,\lambda_2,\ldots,\lambda_{|\Lambda|}$, stopping at the first hyperparameter $\lambda_i$ that does not satisfy the inequality
\begin{equation}
    p_{\lambda_j} \leq \delta.
\end{equation}
It then returns the set 
\begin{equation}
    \hat{\Lambda} = \left\{\lambda_1, \ldots, \lambda_i\right\}
\end{equation}
as the set of reliable hyperparameters. FST can be also extended to consider multiple orderings (see \cite[Algorithm 1]{angelopoulos2021learn}). 
 
An overview of the LTT algorithm can be found in Algorithm \ref{alg::LTT}. The set $\hat{\Lambda}$ produced by LTT satisfies the desired reliability condition (\ref{eq:LTT}) for any FWER-controlling mechanism $\mathcal{A}$\cite{angelopoulos2021learn}.

\begin{algorithm}
\caption{LTT \cite{angelopoulos2021learn} and QLTT [this paper]}
\label{alg::LTT}
\begin{algorithmic}
    \STATE \textbf{Input:} Candidate set $\Lambda$, risk function $R(Z, \lambda)$
    \STATE \textbf{Output:} Set $\hat{\Lambda} \subseteq \Lambda$ of reliable hyperparameters

    \STATE Initialize $\hat{\Lambda} = \emptyset$ 
    
    \IF{method is LTT}
        \STATE Evaluate $p$-values using (\ref{eq:pval}) for all $\lambda \in \Lambda$
    \ELSIF{method is QLTT}
        \STATE Evaluate $p$-values using Proposition \ref{Lemma::interval_to_pvalue} for all $\lambda \in \Lambda$
    \ENDIF

    \STATE Apply an FWER-controlling algorithm: $\hat{\Lambda} = \mathcal{A}(\{p_\lambda\}_{\lambda \in \Lambda}) $

    \RETURN $\hat{\Lambda}$
\end{algorithmic}
\end{algorithm}

\section{Quantile-Based LTT}
\label{sec:QLTT}

While LTT controls the average risk (\ref{Eq:LTT_risk}), many applications require the designer to control other functionals of the test risk $R(Z,\lambda)$. Notably, in engineering applications, one often wishes to control quantiles of the risk $R(Z,\lambda)$, as underlined in Section \ref{sec:motivation}. In this section, we introduce \textit{quantile-based LTT} (QLTT), a method that offers statistical guarantees on a user-defined quantile of the test risk.

To start, define the \textit{$q$-quantile risk} attained by a hyperparameter $\lambda$ as
\begin{equation}
\label{eq:quantile_risk}
    R_q(\lambda) = \inf_{r\geq 0} \left\{\mathrm{Pr}_{\mathcal{Z}}[R(Z,\lambda) \leq r] \geq 1-q  \right\},
\end{equation}
where $q$ is a user-specified outage rate, with $0\leq q\leq 1$, and the probability in (\ref{eq:quantile_risk}) is calculated over the unknown data distribution $p_{\mathcal{Z}}$. The goal of QLTT is to produce a subset of hyperparameters $\hat{\Lambda} \subset \Lambda$ that satisfy the condition
\begin{equation}
\label{eq:QLTT}
    \mathrm{Pr}_{\mathcal{Z}}[R_q(\hat{\lambda}) \leq \alpha \; \text{for all}\; \hat{\lambda} \in \hat{\Lambda}]\geq 1-\delta,
\end{equation}
ensuring that all selected hyperparameters $\hat{\lambda} \in \hat{\Lambda}$ satisfy the constraint $R_q(\hat{\lambda})\leq \alpha$ with probability at least as large as $1-\delta$. To this end, in a manner analogous to LTT, QLTT adopts MHT to test the null hypotheses $\mathcal{H}_\lambda: R_q(\lambda)>\alpha$ for all candidate hyperparameters $\lambda \in \Lambda$. To obtain the $p$-values to be used in Algorithm \ref{alg::LTT}, QLTT leverages the following lemma.

\begin{lemma}[\!\!\!{\cite[Theorem 1]{howard2022sequential}}]
\label{lemma:confidence}
    Given calibration data $\mathcal{Z} = \{Z_i\}_{i=1}^n$, for any $0<\epsilon<1$, the following is a one-sided confidence interval for the $q$-quantile risk $R_q(\lambda)$
    \begin{equation}
        \mathrm{Pr}_{\mathcal{Z}}[ R_q(\lambda) \leq \hat{R}_{q^*}(\lambda, \epsilon)] \geq 1-\epsilon ,
    \end{equation}
    where $\hat{R}_{q^*}(\lambda, \epsilon)$ is the $q^*$-empirical quantile of the risk, i.e., the $\lfloor n(1-q^*)\rfloor$-th smallest element of the set $\left\{ R(Z_j,\lambda)\right\}_{j=1}^n$, with 
    \begin{equation}
        q^* = q- 1.5\sqrt{q(1-q) r_n }  - 0.8  r_n,
    \end{equation}
    and
    \begin{equation}
        r_n = \frac{1.4 \log \log (2.1 n) + \log(10/\epsilon)}{n}.
    \end{equation}
\end{lemma}

Using the duality between confidence intervals and acceptance regions in hypothesis testing \cite{rice2007mathematical}, we obtain a $p$-value for each hypothesis $\mathcal{H}_{\lambda}$ as in the following proposition.

\begin{proposition}
\label{Lemma::interval_to_pvalue}
    Given the quantile estimate $\hat{R}_{q^*}(\lambda, \epsilon)$ as in Lemma \ref{lemma:confidence}, the quantity
    \begin{equation*}
        \hat{p}_\lambda = \inf \{\epsilon \in [0,1]: \hat{R}_{q^*}(\lambda, \epsilon) \geq \alpha \}
    \end{equation*}
    is a $p$-value for the null hypothesis $\mathcal{H}_\lambda$.
\end{proposition}
\begin{proof}
    This follows directly from the standard steps \cite[Chapter 9]{rice2007mathematical}
    \begin{align}
        \mathrm{Pr}_{\mathcal{Z}}[\hat{p}_\lambda \leq u \mid \mathcal{H}_\lambda] & = \mathrm{Pr}_{\mathcal{Z}}[\hat{R}_{q^*}(\lambda,u)< \alpha \mid R_q(\lambda) > \alpha ] \nonumber \\
        &\leq  \mathrm{Pr}_{\mathcal{Z}}[\hat{R}_{q^*}(\lambda,u)<  R_q(\lambda)] \nonumber \\
        &\leq u, 
    \end{align}
    where the last inequality follows from Lemma \ref{lemma:confidence}.
\end{proof}

As summarized in Algorithm \ref{alg::LTT}, having evaluated the $p$-values for the null hypotheses $\mathcal{H}_\lambda$ for all $\lambda \in \Lambda$ using Proposition \ref{Lemma::interval_to_pvalue}, QLTT forms the set $\hat{\Lambda} \subseteq \Lambda$ by applying an FWER-controlling algorithm such as FST. Like LTT, by the definition of FWER, QLTT satisfies the desired reliability condition (\ref{eq:QLTT}).


\section{Simulations and Results}
\label{sec:simulations}

\subsection{Problem Setting}

In this section, we test QLTT on a radio access scheduling problem \cite{de2020radio}. In the setup under study, $N_{B} = 25$ downlink resource blocks need to be allocated to $K=32$ user equipments (UEs), each of which is assigned randomly to one out of four possible quality of service (QoS) classes, with different delay and bit rate requirements \cite{de2020radio}.

Each episode starts by randomly scattering a number $K$ of UEs, each with an initially empty buffer of size 100 packets, in a square area of size 1 $\text{km}^2$. The base station is located at the center of the area. Throughout the episode, the UEs move at random speeds in random rectilinear trajectories as specified in \cite{de2020radio}. Each episode is divided into 10,000 transmission time intervals (TTIs) that last 1 ms. At the beginning of each TTI, new packets get randomly generated for each UE, and added to its buffer. Resource block allocation is carried out for each TTI.

The resource allocation agent is controlled via a hyperparameter $\lambda \in \Lambda$. Specifically, we adopt the learning-based agent proposed by \cite{de2020radio}, whose operation depends on four scalar hyperparameters collected in vector $\lambda = (\lambda_1, \lambda_2, \lambda_3, \lambda_4)$. By definition, the hyperparameters $\lambda_1$, $\lambda_2$, $\lambda_3$ and $\lambda_4$ determine the relative priority of different criteria in the reward model, controlling the channel quality of each UE, the total queue sizes at the UEs, the age of the oldest packet in each UE's buffer, and a fairness measure related to the fraction of resource blocks that were previously allocated to each UE, respectively.

To optimize the hyperparameters $\lambda$, we focus on the delay performance of UEs in the QoS class 1, which has the most demanding QoS requirements. Denote as $D_k(Z,\lambda)$ the packet delay experienced by the $k$th packet for a UE in class 1 and as $K(Z)$ the number of such packets in an episode. Note that $Z$ here refers to the random probabilities involved in a given episode. The risk is then defined as the average delay
\begin{equation}
\label{eq:risk_simulation}
    R(Z,\lambda) = \frac{\sum_{k=1}^{K(Z)}D_k(Z,\lambda)}{K(Z)}.
\end{equation}

We impose that for a fraction of at least $1-q$ of episodes, the risk is not larger than $\alpha = 10$ ms, with probability at least $1-\delta = 0.9$. We compare the results obtained by QLTT to those obtained by LTT, for which the goal is to control the average risk (\ref{eq:risk_simulation}) across the episodes.

We generate the calibration and test data using the Nokia wireless suite \cite{Nokia}. QLTT and LTT were implemented as defined in Algorithm \ref{alg::LTT}, where we used the code provided by \cite{angelopoulos2021learn} for LTT, adopting the Bonferroni correction as the FWER-controlling algorithm. For each run of Algorithm \ref{alg::LTT}, 100 episodes are used as calibration data, and 100 episodes are used as test data. The initial set of 256 candidate hyperparameters $\Lambda$ is obtained starting from the value $\lambda^* = (\lambda_1^*, \lambda_2^*, \lambda_3^*, \lambda_4^*) $ recommended by \cite{de2020radio} by including all combinations of the form $(a_1 \lambda_1^*, a_2 \lambda_2^*, a_3 \lambda_3^*, a_4 \lambda_4^*) $ with $a_1, a_2, a_3, a_4 \in \{1/2, 1, 3/2, 2\}$.

Finally, to select one hyperparameter $\hat{\lambda}$ from the produced set $\hat{\Lambda}$, we opted to maximize the reward signal defined in \cite[Eq. 1]{de2020radio} on the calibration data. Accordingly, the reward for an episode is the negative sum of the number of packets that do not satisfy their respective QoS requirements.

\subsection{Results}

\begin{figure}
    \centering
    \includegraphics[width = \columnwidth]{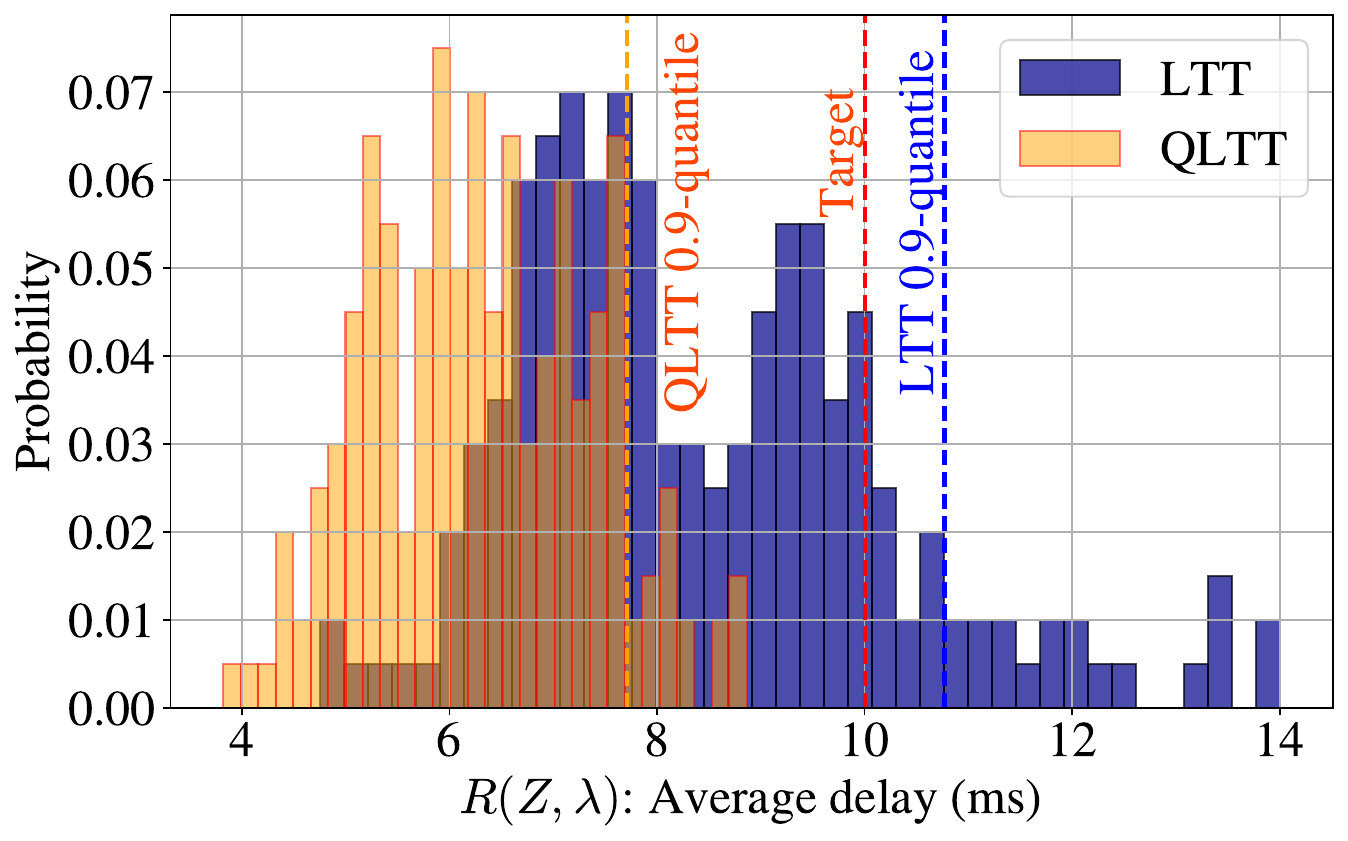}
    \caption{Histogram of the average packet delay $R(Z,\lambda)$ for QoS class 1 for a single run of LTT and QLTT ($q$ = 0.1).}
    \label{fig:histogram}
\end{figure}

To start, Fig. \ref{fig:histogram} shows the histogram of the risk $R(Z,\lambda)$ in (\ref{eq:risk_simulation}), estimated on test data, by using the hyperparameters returned by LTT and QLTT with outage probability $q=0.1$. The histogram is evaluated on a single run of Algorithm \ref{alg::LTT}. Both LTT and QLTT are observed to return average delays for the target QoS class 1 that falls below the 10 ms threshold, satisfying average risk requirements. However, LTT yields a distribution with a heavier upper tail, resulting in a $1-q=0.9$-quantile equal to 10.77 ms. In contrast, the 0.9-quantile of the delay obtained with QLTT is smaller than 10 ms as per the design requirements, validating Proposition \ref{Lemma::interval_to_pvalue}.

\begin{figure}[htbp]
    \centering

    \begin{subfigure}{0.98\columnwidth}
        \centering
        \includegraphics[width=\columnwidth]{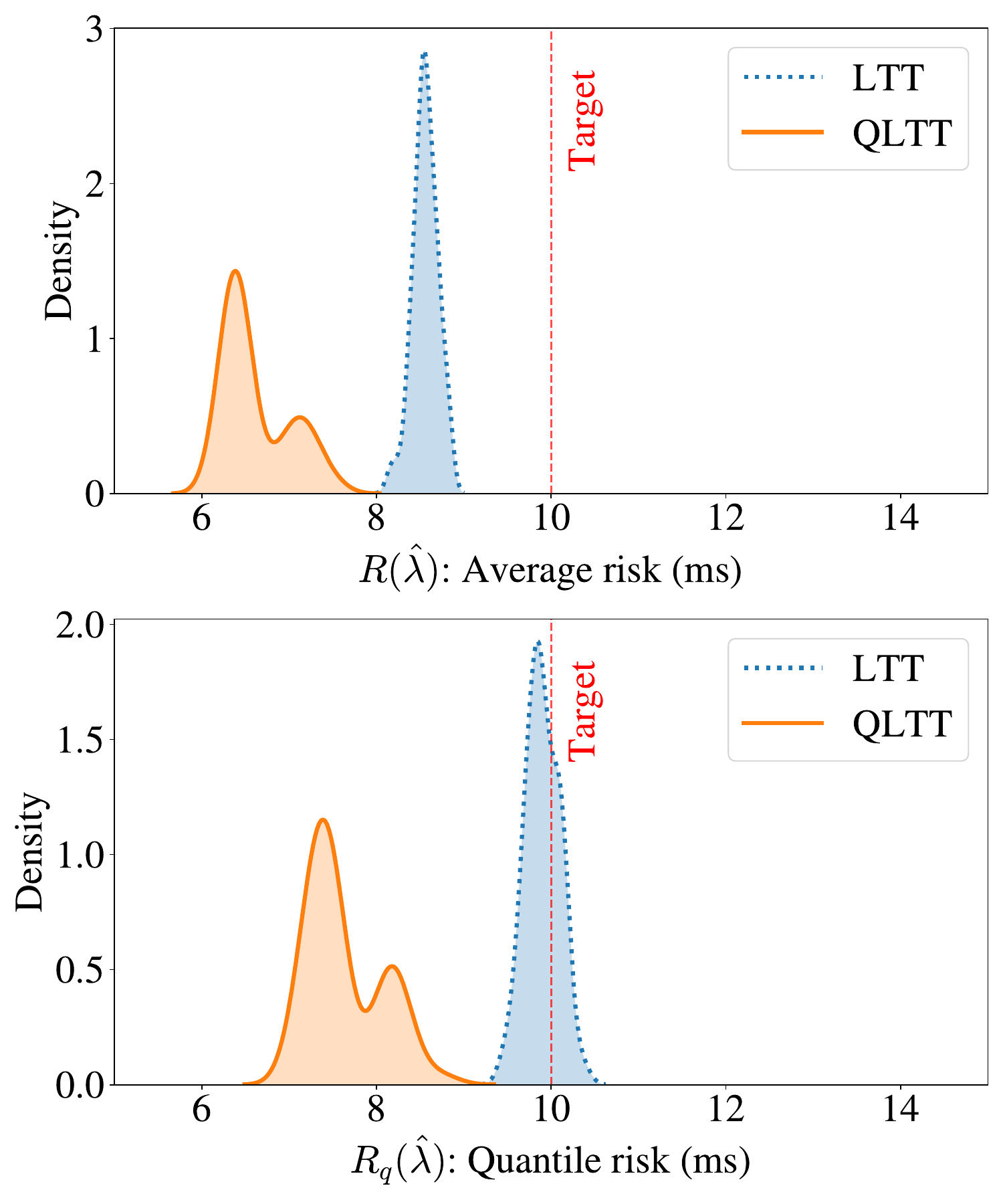}
        \caption{$q$ = 0.2}
        \label{subfig1}
    \end{subfigure}
    \begin{subfigure}{0.98\columnwidth}
        \centering
        \includegraphics[width=\columnwidth]{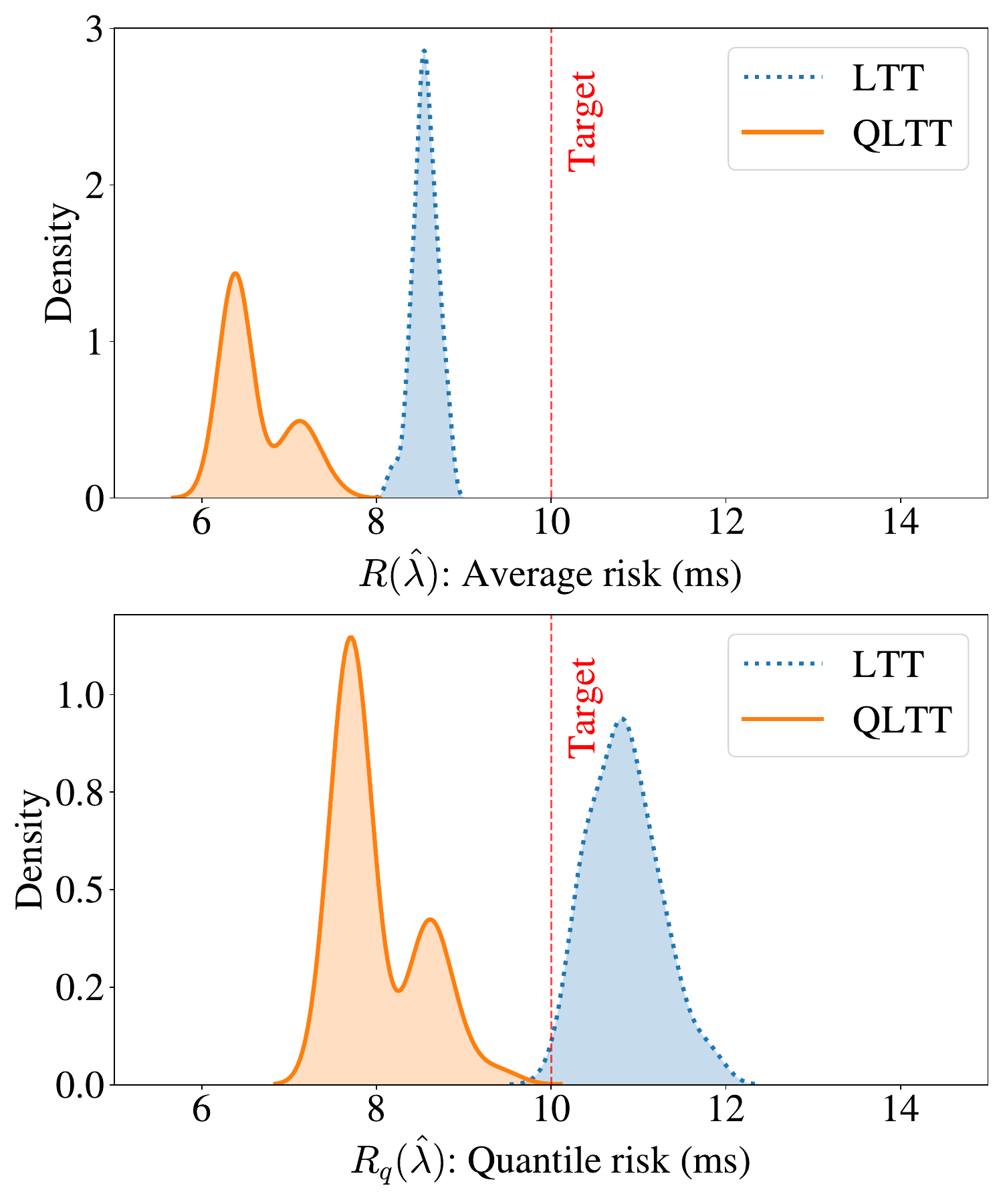}
        \caption{$q$ = 0.1}
        \label{subfig2}
    \end{subfigure}

    \caption{Distribution of the average risk $R(\hat{\lambda})$ and the quantile risk $R_q(\hat{\lambda})$ for LTT and QLTT for outage rates $q = 0.2$ (a) and $q = 0.1$ (b). The distributions are obtained by running Algorithm \ref{alg::LTT} multiple times, obtaining different realizations of hyperparameter $\hat{\lambda}$.}
    \label{fig:violin}
\end{figure}

While Fig. \ref{fig:histogram} was obtained for a single run of Algorithm \ref{alg::LTT}, we now evaluate the distribution of the delay for QoS class 1 across multiple realizations of the calibration data. Fig. \ref{fig:violin} shows the distribution of the average risk $R(\hat{\lambda})$ and the quantile risk $R_q(\hat{\lambda})$ for both LTT and QLTT for two different values of outage rate, namely $q=0.2$ (Fig. \ref{fig:violin}(a)) and $q=0.1$ (Fig. \ref{fig:violin}(b)). Note that the distributions arise here due to the different hyperparameters $\hat{\lambda}$ selected by the calibration procedure when fed distinct realizations of the calibration data. QLTT consistently yields quantile risks below the 10 ms target. In contrast, even though it satisfies the average risk guarantee, LTT returns quantile risks that exceed 10 ms with large probability, particularly as the outage rate $q$ is decreased.

\section{Conclusion}
\label{sec:conclusion}

In this paper, we have addressed risk control in HPO by extending the capabilities of the LTT methodology to include statistical guarantees on quantiles of the risk. This is of particular importance in engineering applications in which designers are often interested in guaranteeing performance levels for given fractions of problem instances. To this end, we have introduced a novel extension, QLTT, which builds upon LTT to offer statistical guarantees for any specified quantile of the risk. The practical efficacy of QLTT was demonstrated through its application to a resource allocation problem, showcasing its ability to address real-world problems. Potential future work can explore further applications of QLTT to other engineering domains, as well as the development of next LTT variants that can offer guarantees on other functionals of distribution of the system variables, such as information measures.

\balance

\bibliographystyle{ieeetr}

\end{document}